\documentclass[twocolumn,showpacs,pra]{revtex4-1}
\usepackage{graphicx}
\usepackage{color}
\usepackage{amsmath}
\usepackage{hyperref}
\usepackage{amssymb}

\newcommand{\ee}{\mathrm{e}}
\newcommand{\ii}{\mathrm{i}}

\newcommand{\br}[1]{\left(#1\right)}

\newcommand{\Ftr}[1]{\mathcal{F}\left[#1\right]}

\newcommand{\phiabb}{\varphi_A}

\begin{document}

\title{Patterned Rydberg excitation and ionisation with a spatial light modulator}

\author{R. M. W. van Bijnen}
\email{rvb@pks.mpg.de}
\altaffiliation[Present address: ]{Max Planck Institute for the Physics of Complex Systems, 01187 Dresden, Germany}

\author{C. Ravensbergen}

\author{D. J. Bakker}

\author{G. J. Dijk}

\author{S. J. J. M. F. Kokkelmans}

\author{E. J. D. Vredenbregt}

\affiliation{Eindhoven University of Technology, P.~O.~Box 513, 5600 MB Eindhoven, The Netherlands}

\date{\today}

\begin{abstract}
We demonstrate the ability to excite atoms at well-defined, programmable locations in a magneto-optical trap, either to the continuum (ionisation), or to a Rydberg state. To this end, excitation laser light is shaped into arbitrary intensity patterns with a spatial light modulator. 
These optical patterns are sensitive to aberrations of the phase of the light field, occuring while traversing the optical beamline. These aberrations are characterised and corrected without observing the actual light field in the vacuum chamber. 
\end{abstract}


\maketitle


Ultracold atomic gases provide a versatile platform for creating and studying well-controlled quantum many-body systems \cite{Bloch08}. 
In particular, highly excited Rydberg atoms offer extremely strong and tunable van der Waals interactions, providing acces to a regime dominated by the interaction energy. The system dynamics is then reduced to that of the internal state of the atoms, while their motion  can be neglected. In this so-called frozen gas limit, Rydberg atoms can for instance be employed in quantum information processing \cite{Saffman09, Saffman10, Wilk10, Isenhower10}, as quantum simulators for interacting spin systems \cite{Weimer08, Weimer10, Ji11, Lesanovsky11, Hague12}, for studying resonant (exitonic) energy transfer \cite{Georg, Ditzhuijzen07, Scholak11, Schoenleber14}, interfacing matter with quantum light \cite{Olmos10, Gorshkov11, Peyronel12, Li13}, and even for emulating wireless networks \cite{Wireless}.


In each of the aforementioned applications a precise control over the locations of the excitations would be greatly beneficial, or even a necessity. 
This control is typically envisioned by restricting the positions of the ground state atoms, e.g., by confining them to an optical lattice and thus enforcing regularly spaced geometries for the excitations to localise on.
In contrast, we consider adding structure to the excitation laser light instead.
Assuming a dense enough underlying atomic density in the frozen gas limit, the possible positions of the Rydberg excitations are then determined solely by the presence of excitation light. In principle, arbitrary geometries can be constructed this way, such as lattice structures exceeding the capabilities and flexibilities of optical lattices. 

\begin{figure}[b]
\includegraphics[width=0.99\columnwidth]{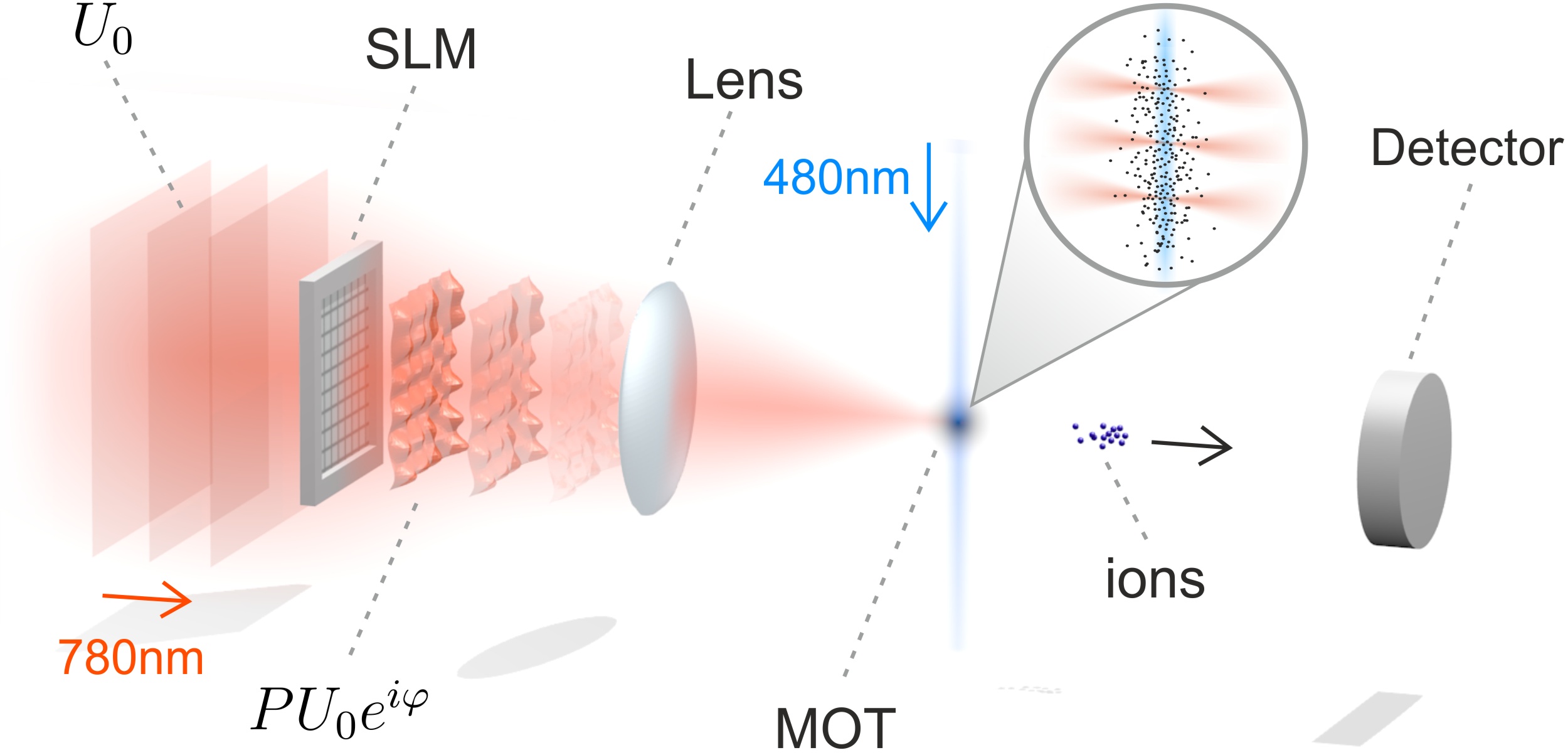}
\caption{Experimental setup, consisting of a red 780 nm and blue 480 nm laser effecting a two-photon transition of atoms in a magneto-optical trap (MOT) to either the continuum (ionisation) or Rydberg level. The SLM imprints a phase onto the 780 nm laser, shaping the intensity into the desired pattern, in this case an array of diffraciton limited spots. Rydberg atoms and ions created at the targeted locations can be accelerated onto a detector which can spatially resolve single ions.\label{FigExcBeamline}}
\end{figure}

There exist several methods of generating arbitrary light fields in cold atom experiments, e.g., time-averaged potentials \cite{Schnelle08, Henderson09}, intensity masking \cite{Scherer07, Brandt11}, or   diffractive optical elements \cite{Dumke02, Newell03, Bakr09, Itah10, Schlosser11}. Here, we employ an active optical element in the form of a spatial light modulator (SLM), which is a commercially available device able to imprint a spatially varying phase onto a laser light field. This way, it is possible to electronically control the intensity of the light field in the focal plane of a lens, as illustrated in Fig. \ref{FigExcBeamline}. 
One great advantage of SLMs is that the intensity patterns are programmatically reconfigurable, without requiring physical alterations to the optical beamline. 
SLMs have already been succesful, for instance, in creating arrays of dipole traps \cite{Bergamini04, He09, Browaeys}, arbitrarily shaped dipole traps \cite{Bruce11, Gaunt12, Gaunt13}, atom guiding \cite{McGloin03, Boyer06, Rhodes06}, and shaping ultracold electron bunches \cite{McCulloch11}. 

Shaping the intensity of the light field through phase modulation requires a high degree of control over the phase of the light field. First, the phase pattern to be imprinted by the SLM needs to be carefully engineered, a task typically performed numerically \cite{Wyrowski88, WyrowskiBook, Pasienski08, Gaunt12, Thesis}.
Second, the quality of the optical light field may be compromised by aberrations introduced by the myriad of components in a typical beamline. This requires a precise characterisation of any phase aberrations experienced by the light field, and subsequent countermeasures need to be taken.
We demonstrate the ability to control the phase of the laser excitation light to sufficiently high degree, and excite atoms at well-defined, programmable locations in a ultracold atomic cloud, either to the continuum (ionisation), or to a Rydberg state.  



Figure \ref{FigExcBeamline} shows the essential part of the experimental setup. 
A magneto-optical trap (MOT), containing a cloud of ultracold ${}^{85}\mathrm{Rb}$ atoms at densities of $10^{16} \mathrm{m}^{-3}$ and temperatures of $0.2 \mathrm{mK}$, is illuminated by two laser beams. A red, 780 nm laser, excites the atoms to an intermediate $5P_{3/2}$ state. A blue, 480 nm laser then ionises the atoms, or excites them to the Rydberg state. A HoloEye Pluto SLM modulates the phase of the 780 nm excitation laser, shaping the intensity pattern in the focal plane of a final lens with a focal length $f = 90\mathrm{mm}$. The blue laser providing the final excitation step is shaped into a thin sheet of light, with a waist diameter of $40\mu \mathrm{m}$ in the $z$-direction, and propagating along the focal plane of the 780 nm laser. The two lasers thus form a 2D excitation volume in the $xy$-plane of the MOT where Rydberg atoms or ions can be created. 

An accelerator extracts the ions and images them onto a detector. In the case of Rydberg atoms the acceleration fields will strip the Rydberg atoms of their valence electron, and the imaging procedure is the same as for ions. During the time of flight the ion pattern does not deform, apart from a global magnification of a factor $46$. The detector, consisting of a double micro-channel plate and phosphor screen, is able to spatially resolve single ions and thus provides a direct measurement of the ion or Rydberg distribution as it was in the MOT, projected onto a plane.

We continue with discussing the phase modulation of the red excitation laser. At the start of the beamline the laser forms a wide beam, to be considered as a plane wave with amplitude $U_0$. The SLM imprints not only a phase $\varphi$ onto the light field, it additionally acts as an aperture with a pupil function $P$. This function is valued $1$ inside the active area, and $0$ outside, truncating the beam to a rectangle with the dimensions $S_x \times S_y = 15 \mathrm{mm} \times 8 \mathrm{mm}$ of the SLM. In scalar diffraction theory and under the Fresnel approximation \cite{Goodman}, the amplitude of the light field $U_f$ in the focal plane follows from a Fourier transform of the light field immediately behind the SLM,
\begin{equation}\label{EqFourier}
|U_f(x', y')| = \frac{1}{\lambda f}\left|\Ftr{P U_0 \exp(\mathrm{i} \varphi(x, y))}\br{\frac{x'}{\lambda f}, \frac{y'}{\lambda f}}\right|,
\end{equation}
where $x', y'$ are coordinates in the focal plane, and $\Ftr{\cdot}(u,v)$ denotes a 2-dimensional Fourier transform of the function in square brackets, evaluated at spatial frequencies $u, v$.

The degree of freedom provided by $\varphi$ can thus be used to modulate the intensity in the focal plane. 
In particular, we focus on creating arrays of diffraction limited spots. It is instructive to first consider the trivial case of a single spot, by choosing $\varphi = 0$. Writing $u = x'/\lambda f$, and $v = y'/\lambda f$, we recover the well known Airy spot:
\begin{equation}\label{EqPspot}
\Ftr{P U_0}(u, v) = U_0 \frac{S_xS_y}{\lambda f} \ \mathrm{sinc}(S_x u) \ \mathrm{sinc}(S_y v),
\end{equation}
where $\mathrm{sinc}(x) = \sin(\pi x) / \pi x$. The size of the central spot $\Delta x \times \Delta y$ between the first minima follows from the diffraction limit
\begin{equation}\label{EqDiffLimit}
\Delta x \times \Delta y = \frac{2\lambda f}{S_x}\times \frac{2\lambda f }{S_y},
\end{equation}
which determines the size of the smallest possible spot we can make in our setup.

\begin{figure}[t]
\includegraphics[width=0.99\columnwidth]{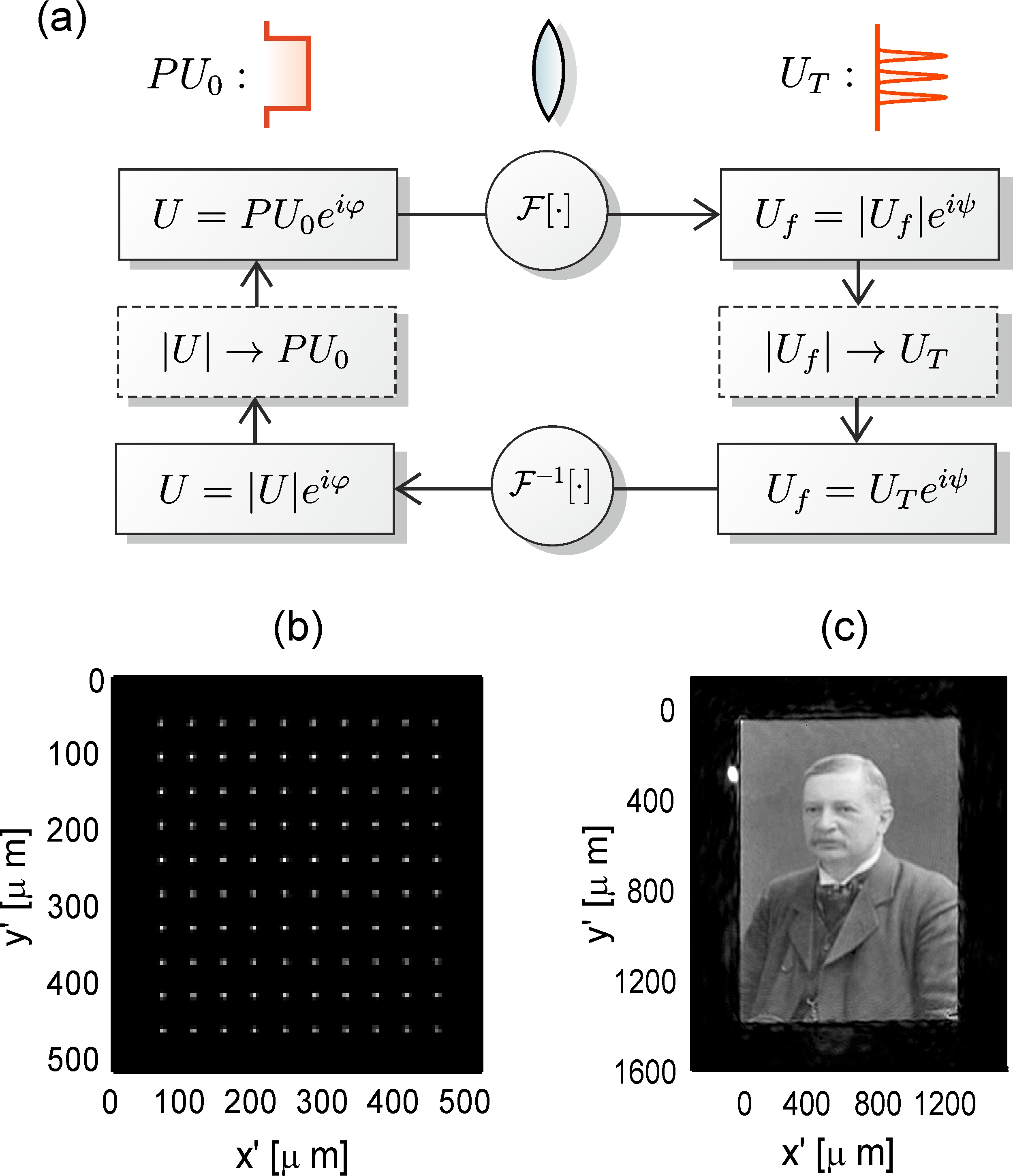}
\caption{(a) Iterative Fourier transform algorithm (IFTA) used for calculating the SLM phase patterns, consisting of a series of virtual propagations (circles) between the SLM plane and the focal plane, while applying constraints (dashed boxes). (b) Camera image of the 780 nm laser beam, shaped into an array of diffraction limited spots with the IFTA algorithm. (c) Portrait of Johannes Rydberg, made out of 780nm laser light, with a modified IFTA algorithm.\label{FigIFTA}}
\end{figure}
In order to compute phases $\varphi$ for complicated intensity patterns such as spot arrays, we employ an iterative Fourier transform algorithm (IFTA) \cite{Wyrowski88, WyrowskiBook, Thesis}, as depicted schematically in Fig. \ref{FigIFTA}(a). Starting from an initial guess for the phase, the algorithm consists of a series of virtual propagations of the light field back and forth between the SLM and focal planes. In each plane, constraints to the amplitude of the light field are applied. In the SLM plane, the amplitude is constrained to that of the input field, $U_0$. In the focal plane, the constraint is formed by the target amplitude, denoted by $U_T$. 
These steps can be viewed as a series of projections between two subsets of $\mathbb{C}^2$, where in each subset one of the two constraints is satisfied \cite{Levi84, Luepken92}. The algorithm converges to a phase $\varphi$ which minimizes the distance between both sets, i.e., satisfying both constraints as well as possible.

In practice, we are only interested in the amplitude of the light field in a small window $W$ in the focal plane, with associated pupil function $P_W$. Restricting the constraint to the window, i.e. setting $|U_f| \to P_W U_T + (1 - P_W) |U_f|$ in Fig. \ref{FigIFTA}(a), provides the algorithm with a greater degree of freedom for the phase pattern, resulting in better convergence and better quality of the final solution \cite{Akahori86}. 
Fig. \ref{FigIFTA}(b) shows a direct camera image of the laser light, which is shaped into an array of diffraction limited spots using the methodology described above. Fig. \ref{FigIFTA}(c) shows a direct camera image of a complicated intensity pattern, demonstrating the  enormous degree of control over the light field that can be effected by the SLM. The computation of the corresponding phase pattern is described elsewhere \cite{Thesis, FeedbackPaper}.


However, shaping the laser intensity by phase modulation is sensitive to aberrations of the phase of the light field, which are inevitably introduced by imperfections in the various optical components in the beamline. 
We model the phase profile $\varphi$ of the light in a plane immediately behind the last optical element as
\begin{equation}\label{EqPhiabb}
\varphi(x, y) = \varphi_0(x, y) + \varphi_L(x, y) + \phiabb(x, y),
\end{equation}
where $\varphi_0$ is the phase imprinted on the light field by the SLM. The phase $\varphi_L$ is the lens phase of an idealised lens focusing at the MOT and effectively performing the Fourier transform. In writing the first term on the rhs. of Eq. (\ref{EqPhiabb}), we have assumed that a ray of light starting at point $(x,y)$ in the SLM plane, remains close to the transverse coordinates $(x,y)$ during its traversal of the beamline \cite{NoteApproximation1}.
The phase $\phiabb$ contains all aberrations, i.e., any deviation of the actual phase from the required phase.
Clearly, the aberrations $\phiabb$ can be corrected by the SLM by subtracting $\phiabb(x, y)$ from $\varphi_0$. 

\begin{figure}\begin{center}
\includegraphics[width=0.99\columnwidth]{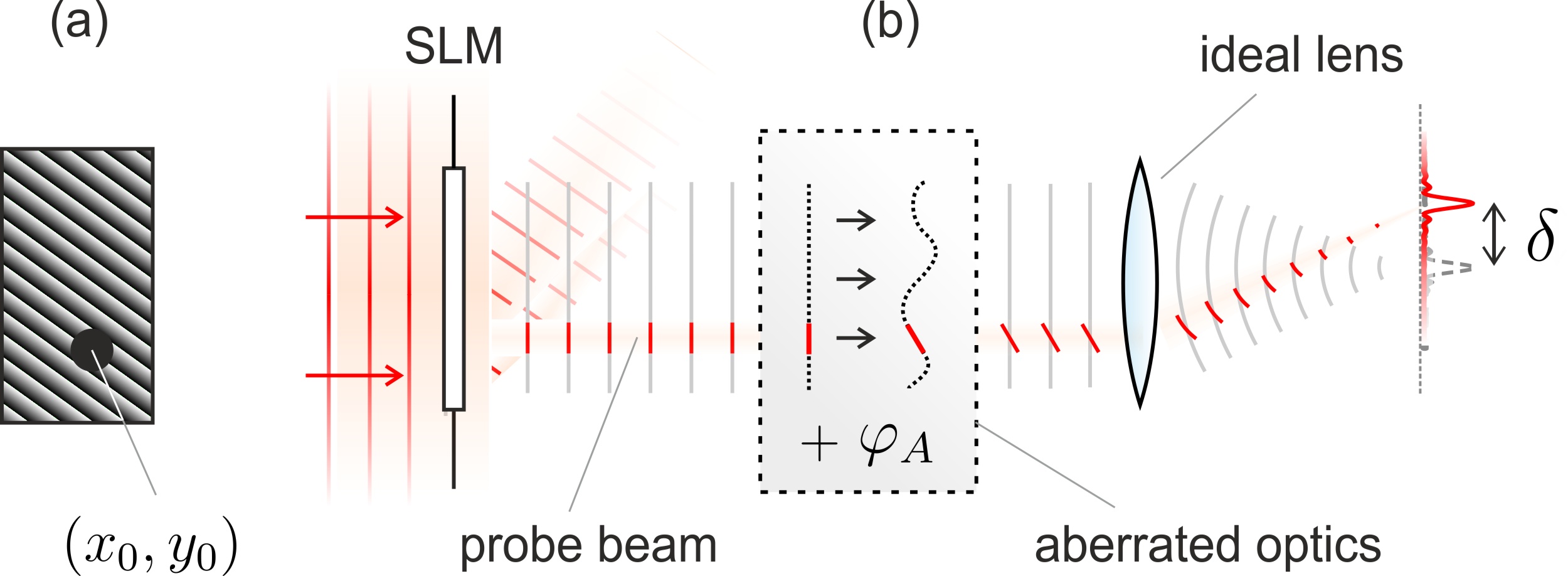}
\caption{Method for sampling the phase aberration $\phiabb$ of the optical system. (a) SLM phase pattern, used for (b) sending a small probe beam through the beamline, which experiences only a small, approximately linear part of $\phiabb(x, y)$. The gradient of $\phiabb$ can then be recovered from the displacement $\delta$ of the focal spot. In practice, the lens is part of the optical system, but for conceptual simplicity it is drawn separately. 
\label{FigAberrProbe}}
\end{center}\end{figure}

A key step is the (nontrivial) determination of the exact phase error inside the vacuum chamber. We resort to a method reminiscent of the Shack-Hartmann procedure \cite{OpticalShopTesting,Bowman10}.
We divide the phase pattern on the SLM into two parts, as shown in Fig. \ref{FigAberrProbe}(a). A small circular area of the SLM is selected, and imprinted with a constant phase. 
The remaining area of the SLM \textit{outside} the circular region is imprinted with a linear phase pattern which causes all the light incident on it to be deflected at a large angle, such that it no longer hits the MOT. Therefore, the excitation laser impinging on the MOT is reduced to a single narrow beam, originating from the small circular area on the SLM. One can think of the phase pattern as if modifying the original pupil function $P$ to an effective pupil function $P'$ corresponding to the small circular area.

As the narrow beam passes through the optical system, it samples only a small part of the total phase aberration, in a small area around the coordinates $(x_0, y_0)$ [see Fig. \ref{FigAberrProbe}(b)]. To a first approximation the phase aberration in this small area, as experienced by our probe beam, is linear in $x$ and $y$.
The effect of a purely linear phase is to move the intensity pattern in the focal plane:
\begin{equation}\label{EqFtrLinear}
\Ftr{P' U_0\ \ee^{2 \pi\ii (a x + by)}}(u, v) = \Ftr{P' U_0}(u - a, v - b).
\end{equation}
Our phase pattern thus produces a single spot, whose size and shape are determined by the pupil function $P'$, whereas its displacement is proportional to the gradient of the phase aberration,  with $2 \pi a = \partial \phiabb / \partial x,$ and $2 \pi b = \partial \phiabb / \partial y$ in Eq. (\ref{EqFtrLinear}). 
The position of the spot in the focal plane is detected by measuring the displacement of the ion spot on the detector, allowing us to determine the local gradient of the phase aberration. By varying the coordinates $(x_0, y_0)$ we can sample $\nabla \phiabb$ in its entirety. By subsequently fitting the gradient data with a smooth polynomial composed of rectangular Zernike polynomials \cite{Mahajan07}, we reconstruct the total phase aberration $\phiabb$.

\begin{figure}[b]\begin{center}
\includegraphics[width=\columnwidth]{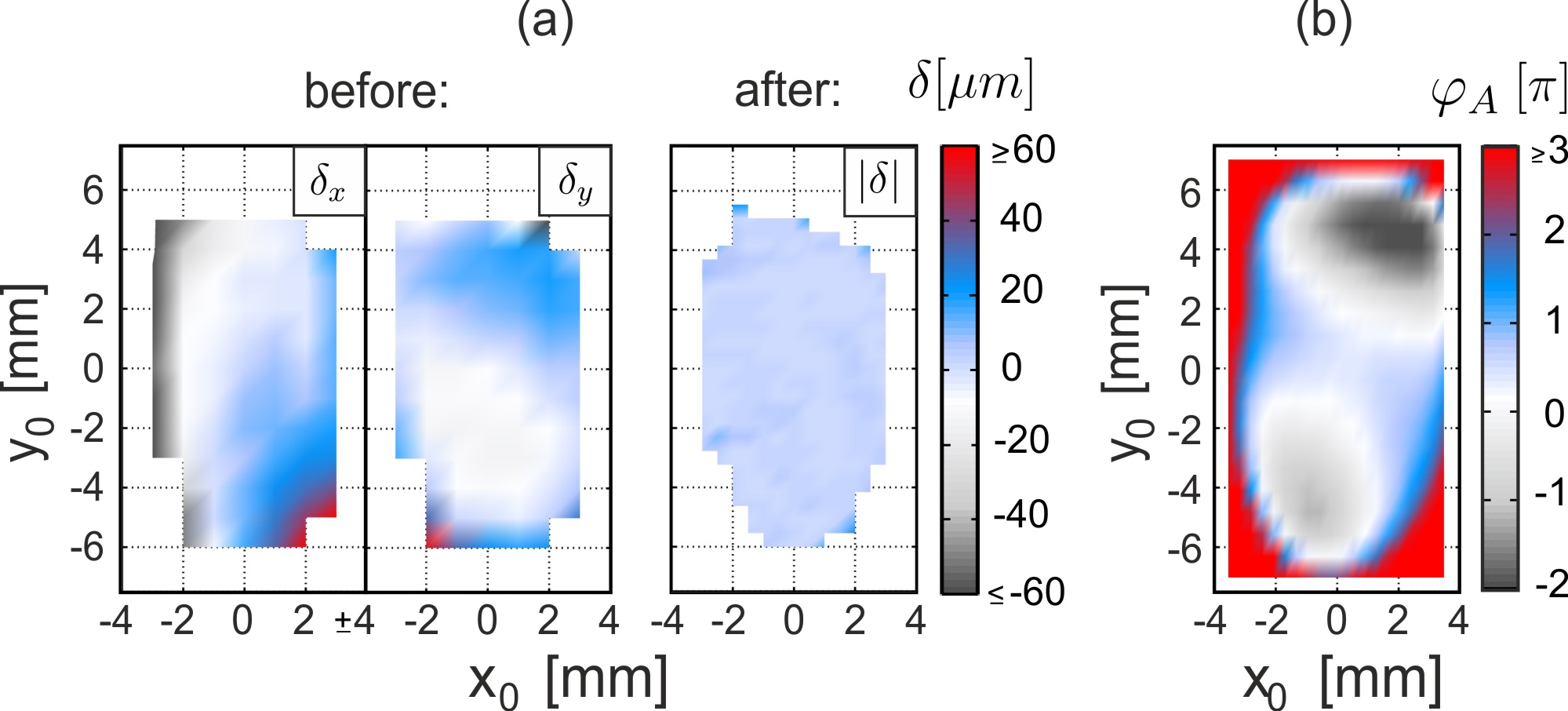}
\caption{(a) Measured spot displacement $\delta$ before and after aberration correction, showing significant flattening of the wavefront. The probe beam is blocked near the edges, hence no data is shown there. (b) Phase aberration $\phiabb$ as subtracted from the wavefront.
\label{FigIonAberrCorr}}
\end{center}\end{figure}

Fig. \ref{FigIonAberrCorr}(a) shows the measured ion spot displacements in the $x$ and $y$ directions, denoted by $\delta_x, \delta_y$, respectively, as a function of the sampling coordinates $(x_0, y_0)$ on the surface of the SLM. The sampling beam diameter was set at $1 \mathrm{mm}$, and a total of $20\times20 = 400$ measurements are taken. There is a clear dependence of the measured spot displacement on the sampling coordinates. The corresponding root-mean-square (RMS) wavefront error is more than a wavelength, which can be classified as highly aberrated.
Near the edges of the sampling area no ion signal was visible, and hence no data points are shown. Light emanating from these areas of the SLM is obstructed by the accelerator structure before it reaches the MOT.
The total usable area of the SLM is determined to (approximately) range between $-5 \mathrm{mm} < x_0 < +5\mathrm{mm}$, and $-3\mathrm{mm} < y_0 < +3\mathrm{mm}$.

The reconstructed phase aberration $\phiabb$, as shown in Fig. \ref{FigIonAberrCorr}(b), is then subtracted from the light field by the SLM. In practice, the procedure is iterated $\sim 5$ times, since the path of the probe beam through the optical system is slightly altered by imprinting a phase with the SLM.
The final spot displacements after iteratively correcting the phase aberrations (see Fig. \ref{FigIonAberrCorr}) are significantly reduced, and the aberrations have been removed to a large extent. The RMS wavefront error associated with the final displacements is approximately $\lambda / 18.7$, suggesting a performance at the diffraction limit ($< \lambda / 14$ \cite{BornWolf}).



\begin{figure}[t]\begin{center}
\includegraphics[width=0.99\columnwidth]{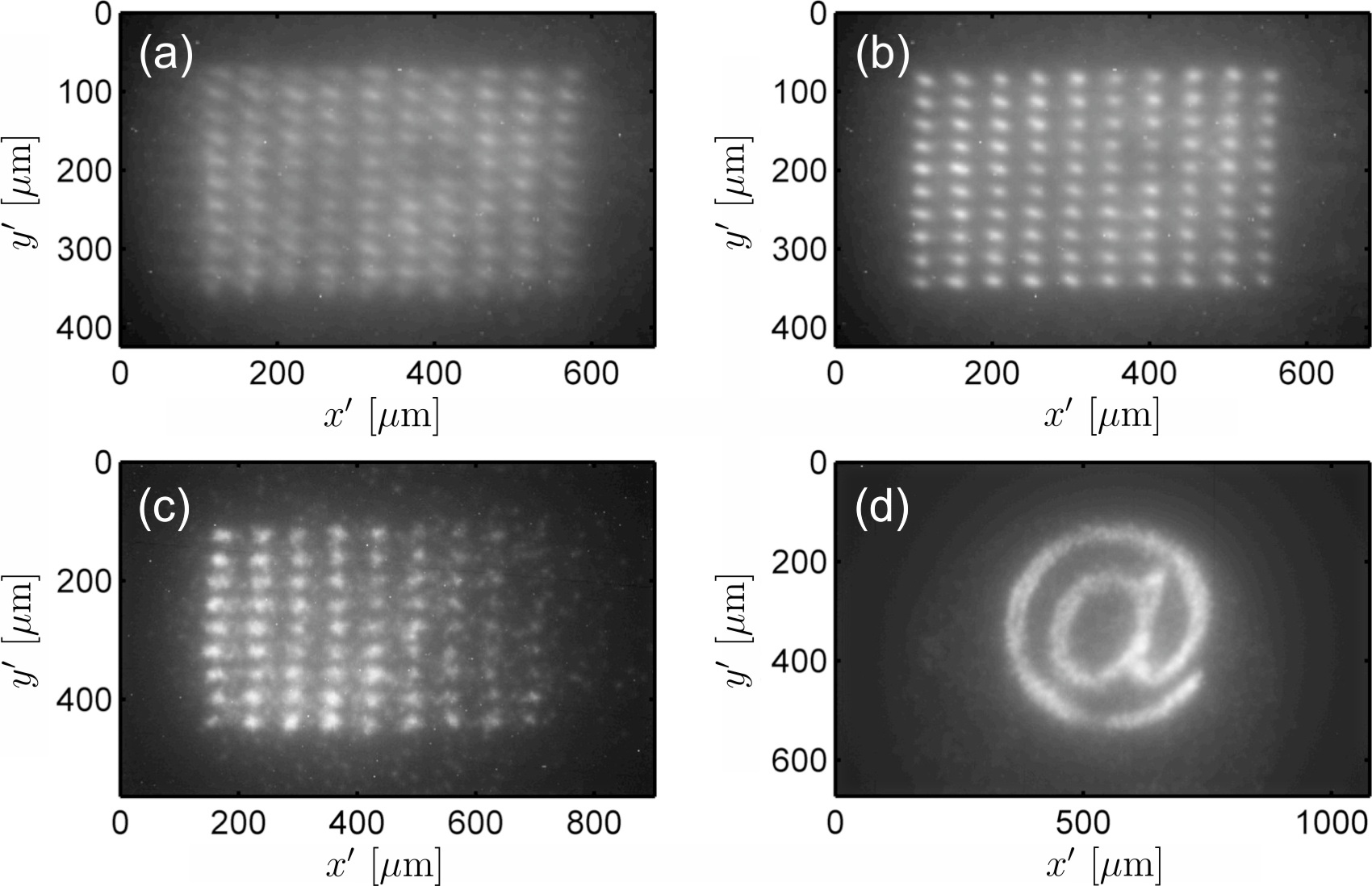}
\caption{Ion detector signal. (a): $10 \times 10$ array of spots, no aberration correction, excitation to the continuum (ionisation). (b): Identical settings as in (a), but after aberration correction. (c): $10\times10$ array of spots, atoms excited to the $82D$ Rydberg state. (d): Complex pattern demonstrating the ability of the SLM to create arbitrary intensity patterns (ionisation). The spatial coordinates are calculated back to positions $x', y'$ in the MOT from where the detected ions originated.\label{FigIons}}
\end{center}\end{figure}

%

After correcting the phase aberration we proceed by creating intensity patterns consisting of arrays of spots. For a single experimental shot we illuminate the cloud  below saturation intensity for $20 \mu \mathrm{s}$, after which the ions are extracted. The detector is exposed for $100\mathrm{s}$ at a $10\mathrm{Hz}$ repetition rate.
Fig. \ref{FigIons}(a) shows the resulting ion detector image with a $10 \times 10$ spot array, measured \textit{before} the aberration correction was applied. Fig. \ref{FigIons}(b) shows the same spot pattern, \textit{after} applying the aberration correction while keeping all other parameters the same, showing a significant improvement in pattern quality. 
The spacing between the spots is approximately $60 \mu \mathrm{m}$ in the $x$-direction, and $30 \mu \mathrm{m}$ in the $y$-direction, with a spot size which would correspond to a diffraction limit of $24 \mu \mathrm{m} \times 15 \mu \mathrm{m}$.  This agrees well with the theoretical diffraction limit from Eq. (\ref{EqDiffLimit}), taking into account the reduced aperture due to beam clipping. Smaller spacings are possible, we are able to resolve individual spots down to a spacing of $25 \mu \mathrm{m} \times 14 \mu \mathrm{m}$. Fig. \ref{FigIons}(c) shows an array of spots, with the atoms in the target locations excited to the Rydberg 82D state. The spacing between the spots is approximately $70 \mu \mathrm{m}$ in the $x$-direction, and $35 \mu \mathrm{m}$ in the $y$-direction. A distinct intensity gradient is visible across the spot pattern, due to density variations of the atomic gas.
Finally, Fig. \ref{FigIons}(d) further demonstrates the capability of the SLM to produce arbitrary intensity patterns in the MOT.

%
In conclusion, we have demonstrated the ability to excite atoms to the Rydberg state or the continuum at well defined, programmable locations in an ultracold atomic gas. We employed an SLM  for delivering high quality optical patterns inside a vacuum chamber, controlling the phase of the light field to well within a wavelength, despite a highly aberrated optical beamline. The aberrations are characterised by observing the ion signal resulting from small probe beams, that are sent by the SLM along different paths through the optical beamline. The SLM could then be used to subtract the phase aberration from the light field, removing the aberrations.




\bibliography{ReferencesAberrations}

\begin{thebibliography}{52}
\bibitem [{Not()}]{NoteApproximation1}%
  \BibitemOpen
  \href@noop {} {}\bibinfo {note} {The validity of this assumption depends also
  on the SLM phase pattern $\varphi_0$, which can affect the path of light
  beams, and it depends on the length of the beamline. We have verified that
  the approximation holds for our experimental conditions.}\BibitemShut {Stop}%
\end{thebibliography}

\end{document}